\documentclass[pmlr]{jmlr}

\RequirePackage{graphicx}
\usepackage{booktabs}
\usepackage{longtable}
\usepackage{amsmath}
\usepackage{bbm}

\makeatletter
\def\set@curr@file#1{\def\@curr@file{#1}} 
\makeatother
\usepackage[load-configurations=version-1]{siunitx} 

\theorembodyfont{\upshape}
\theoremheaderfont{\scshape}
\theorempostheader{:}
\theoremsep{\newline}

\jmlrvolume{}
\jmlryear{}
\jmlrworkshop{}

\title[]{Osteoporosis Prediction from Hand X-ray Images Using Segmentation-for-Classification and Self-Supervised Learning}

\author{\Name{Ung Hwang}
       \Email{hwangung@gmail.com}\\ 
       \addr Department of Electronic Engineering\\
       Hanyang University\\
       Seoul, Korea
       \AND
       \Name{Chang-Hun Lee}
       \Email{drlch79@hanyang.ac.kr}\\ 
       \addr Department of Orthopedic Surgery\\
       Hanyang University\\
       Seoul, Korea
       \AND
       \Name{Kijung Yoon}
       \Email{kiyoon@hanyang.ac.kr}\\ 
       \addr Department of Electronic Engineering\\
       \addr Department of Artificial Intelligence\\
       Hanyang University\\
       Seoul, Korea}

\begin{document}

\maketitle

\begin{abstract}
  Osteoporosis is a widespread and chronic metabolic bone disease that often remains undiagnosed and untreated due to limited access to bone mineral density (BMD) tests like Dual-energy X-ray absorptiometry (DXA). In response to this challenge, current advancements are pivoting towards detecting osteoporosis by examining alternative indicators from peripheral bone areas, with the goal of increasing screening rates without added expenses or time. In this paper, we present a method to predict osteoporosis using hand and wrist X-ray images, which are both widely accessible and affordable, though their link to DXA-based data is not thoroughly explored. We employ a sophisticated image segmentation model that utilizes a mixture of probabilistic U-Net decoders, specifically designed to capture predictive uncertainty in the segmentation of the ulna, radius, and metacarpal bones. This model is formulated as an optimal transport (OT) problem, enabling it to handle the inherent uncertainties in image segmentation more effectively. Further, we adopt a self-supervised learning (SSL) approach to extract meaningful representations without the need for explicit labels, and move on to classify osteoporosis in a supervised manner. Our method is evaluated on a dataset with 192 individuals, cross-referencing their verified osteoporosis conditions against the standard DXA test. With a notable classification score, this integration of uncertainty-aware segmentation and self-supervised learning represents a pioneering effort in leveraging vision-based techniques for the early detection of osteoporosis from peripheral skeletal sites.
\end{abstract}

\section{Introduction}
\label{sec:intro}
Osteoporosis is a common bone ailment characterized by reduced bone mineral density (BMD) or bone mass loss, leading to bones becoming fracture-prone and structurally compromised. Given its prevalence and far-reaching impact, there is a pressing need for preemptive risk assessment, early diagnosis, and effective preventive actions. While computed tomography (CT) and magnetic resonance imaging (MRI) have demonstrated potential in BMD estimation and osteoporosis screening \citep{chou2017vertebral,pickhardt2013opportunistic,gausden2017opportunistic}, their clinical use is limited due to concerns about radiation exposure and associated costs. At present, the dual energy X-ray absorptiometry (DXA) is recognized as a standard and reliable instrument for osteoporosis detection and BMD analysis. Nonetheless, DXA comes with its own set of challenges: the proficiency of the operator can influence the results, the patient's posture during the test can skew accuracy, and in obese patients with significant fat mass, BMD might be overestimated \citep{watts2004fundamentals,messina2015prevalence}.

As a solution, recent studies have highlighted the utility of X-ray images, routinely acquired in clinical settings, as an effective means for gathering comprehensive data. This approach can be particularly useful for osteoporosis detection \citep{yamamoto2020deep,yasaka2020prediction,hsieh2021automated,jang2021prediction,ho2021application}, leveraging the widespread availability of radiographs from individuals who have not been specifically screened for the condition. By eliminating the need for expensive DXA equipment, this strategy presents a more affordable option and is preferable for patients due to its considerably lower radiation exposure compared to DXA scans. Among the various clinical approaches, the 2nd metacarpal cortical index (2MCI) in hand X-ray images has gained attention as a promising biomarker for osteoporosis screening \citep{schreiber2017simple,patel20202nd}, with the cortical thickness and cancellous bone tissue porosity of the metacarpal bone playing a crucial role in disease identification. Despite its promise, this method faces obstacles, such as the time-intensive manual measurement of metacarpal dimensions and its failure to encompass all disease characteristics. A viable alternative is automated radiographic understanding and analysis, which can offer valuable insights to orthopedists during the diagnostic process.

\begin{figure}[t]
\includegraphics[width=\linewidth]{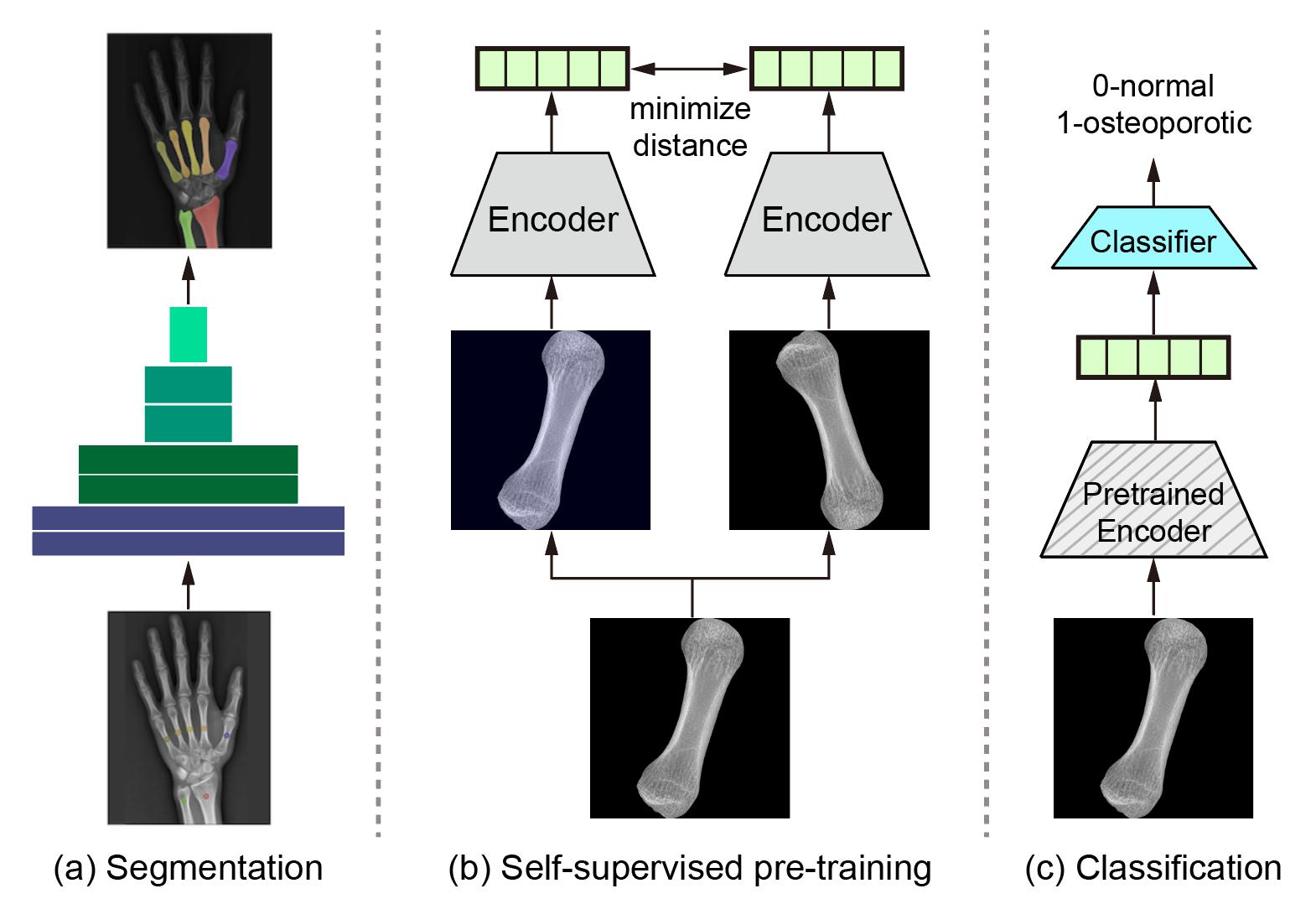}
  \caption{Illustration of framing a hand X-ray osteoporosis screening problem through (a) segmentation, (b) self-supervised pre-training, and (c) downstream classification task.}
  \label{fig:overview}
\end{figure}

To address the research gap between traditional manual radiographic analysis and the recent strides in deep learning that have reached diagnostic levels comparable to those of medical experts in various fields \citep{huang2023self}, we introduce a novel automated approach for osteoporosis screening (Figure \ref{fig:overview}). In these procedures, we innovate by redefining the task of medical image classification as a segmentation (Figure \ref{fig:overview}a), thereby obviating the need for manual calculations of the 2MCI. The first step in our approach involves curating the training dataset by identifying and isolating critical bones in hand and wrist radiographs, with a particular focus on the metacarpals for osteoporosis screening. This task is accomplished through a dedicated U-Net architecture \citep{ronneberger2015u,kohl2018probabilistic,baumgartner2019phiseg} designed to effectively captures the multimodal aspects of predictive uncertainty in semantic segmentation. It is akin to the routine practices of orthopedists who depend on quantitative assessments derived from segmentation masks. Following the segmentation-based classification reformulation, our approach employs a self-supervised learning (SSL) framework (Figure \ref{fig:overview}b) for the initial pre-training phase of our classification networks (Figure \ref{fig:overview}c). This framework enables the models to discern nuanced, yet critical features without the dependency on explicitly labeled data, thereby improving the models' generalization capabilities, especially in the context of limited data availability, as demonstrated in our study with 192 subjects. Furthermore, we enhance the efficiency of our self-supervised learning by incorporating a customized multi-crop augmentation strategy \citep{caron2020unsupervised}, specifically adapted for hand X-ray images, to optimize the learning process.

This paper explores how easily accessible, yet unlabeled, hand radiographic data can be leveraged to detect osteoporosis on real-world medical settings. Our contributions are threefold:
\begin{itemize}
    \item We introduce a novel automated approach for osteoporosis screening that utilizes peripheral radiographic images instead of central DXA measurements, leveraging a custom U-Net architecture to eliminate the need for traditional 2MCI calculations. This shift could significantly simplify and enhance the accessibility of osteoporosis screenings.

    \item We explore a SSL framework focused on segments of hand and wrist bones. This method allows our models to effectively distill critical features prior to osteoporosis prediction, demonstrating enhanced outcomes compared to conventional supervised learning approaches.

    \item Our research boosts model training efficiency through a specialized multi-crop augmentation strategy specifically designed for hand X-ray images, optimizing the learning process and improving model performance in scenarios with limited data availability.

\end{itemize}

\section{Related Work}
\paragraph{Osteoporosis Prediction in Radiographic Data.}

Research in osteoporosis prediction has continuously embraced a variety of radiographic imaging techniques and computational models to enhance diagnostic accuracy. For example, \citet{yamamoto2020deep} and \citet{jang2021prediction} have expanded the use of deep learning models on hip radiographs, frequently integrating clinical data to boost the models' diagnostic effectiveness. Similarly, \citet{hsieh2021automated} illustrated the effectiveness of traditional radiographic scans of the pelvis and lumbar spine in predicting bone mineral density (BMD) and assessing fracture risk, alongside automated tools. \citet{ho2021application} also utilized plain pelvis X-rays to introduce the DeepDXA model, which uses convolution-based regression to predict BMD, highlighting the potential for broad screening applications with standard radiographic images.

Recent studies have shifted towards leveraging more readily accessible imaging types; for instance, \citet{wang2022lumbar} employed chest X-rays and developed a multi-ROI deep model with a transformer encoder to precisely estimate BMD, while \citet{sebro2022machine} explored the use of computed tomography in analyzing wrist and forearm CT scans for opportunistic screening, revealing positive correlations between CT attenuations of the wrist/forearm bones and DXA measurements. In a slightly different approach, \citet{zheng2016bone} applied texture-based descriptors to distinguish between healthy and osteoporotic subjects, supporting the role of 2D texture analysis in detecting changes in trabecular bone microarchitecture. Our work distinguishes itself by focusing on hand radiographs, leveraging these peripheral yet commonly available radiographic data. We enhance our approach by applying self-supervised learning on the hand X-rays, aiming to encourage the model to learn features within the cluster of articulating bones in each hand and wrist, which are likely directly correlated with osteoporotic labels.

\paragraph{Medical Image Classification as Segmentation.}
The recent trend in medical image analysis has seen an innovative convergence of segmentation techniques with classification tasks, considerably increasing diagnostic accuracy and model robustness. This synthesis allows for the leveraging of spatial and structural features inherent in segmentation tasks to improve the classification of various medical conditions. Early endeavors, exemplified in \citet{wong2018building}, introduced a curriculum learning-inspired approach where features from segmentation networks are utilized to facilitate classifying complex structures, such as brain tumors and cardiac levels, by initially training networks to understand simpler shape and structural concepts through segmentation. Further research, as discussed in \citet{heker2020joint} and \citet{mojab2021classification}, continued to expand on this foundation by blending transfer and joint learning techniques to optimize feature extraction and model robustness in medical imaging tasks, such as liver lesion segmentation and classification. These studies underline the effectiveness of using segmentation as a preliminary step to classification, achieving notably better results than traditional multi-task architectures. In \citet{gare2021dense}, authors took a different approach by employing pre-trained segmentation models for diagnostic classification, termed reverse-transfer learning, which highlighted the advantages of dense versus sparse segmentation labeling and reduced false positive rates in lung ultrasound analysis. Finally, \citet{saab2022reducing} addressed the reliance of neural networks on spurious features by increasing spatial specificity through segmentation. By providing more precise location data of abnormalities, the models achieved greater robustness against misleading features, which was particularly important in tasks like pneumothorax and melanoma classification. Each of these studies highlights a unique dimension of how segmentation can enrich classification tasks in medical imaging, from increasing data richness to improving model reliability and reducing the need for extensive annotated datasets. Our work builds upon these advancements by specifically applying segmentation to derive features from the ulna, radius, and five metacarpals within a self-supervised learning framework for hand and wrist radiographs.

\section{Methods}
\label{sec:methods}
We begin with the hypothesis that directly employing raw X-ray images may not serve as the ideal input for our residual neural networks for classification due to their large dimensions, which surpasses the handling capacity of standard image classification frameworks. Moreover, merely resizing these images could pose the risk of losing essential features such as bone texture and microarchitecture \citep{lespessailles2008clinical,zheng2016bone,sebro2022machine}, vital for making precise osteoporosis predictions. As an alternative, instead of using arbitrary small image patches, we propose using images based on specific bones, obtained through segmentation of the targeted areas. This approach of segment-specific analysis is backed by several pieces of research demonstrating the benefits of incorporating segmentation methods into classification tasks \citep{gare2021dense,gare2022w,heker2020joint,mojab2021classification,wong2018building,saab2022reducing,hooper2024case}. Our approach also aligns with findings from earlier research indicating a significant correlation between the 2MCI, which can be derived from segmented inputs, and BMD measurements \citep{schreiber2017simple,patel20202nd}. Consequently, we opt to isolate individual bones in hand and wrist radiographic images to enhance the specificity of our analysis.

In this section, we outline the process for creating segmentation masks for the bones of interest (Sec. \ref{sec:mask}), introduce multi-crop data augmentation methods (Sec. \ref{sec:aug}), and detail our approach to applying self-supervised pre-training to segmented image inputs (Sec. \ref{sec:ssl}) followed by supervised classification task (Sec. \ref{sec:finetuning}).

\subsection{Segmentation Mask for Target Bones}
\label{sec:mask}

\subsubsection{Problem Setup}
\begin{figure}[t]
\includegraphics[width=\linewidth]{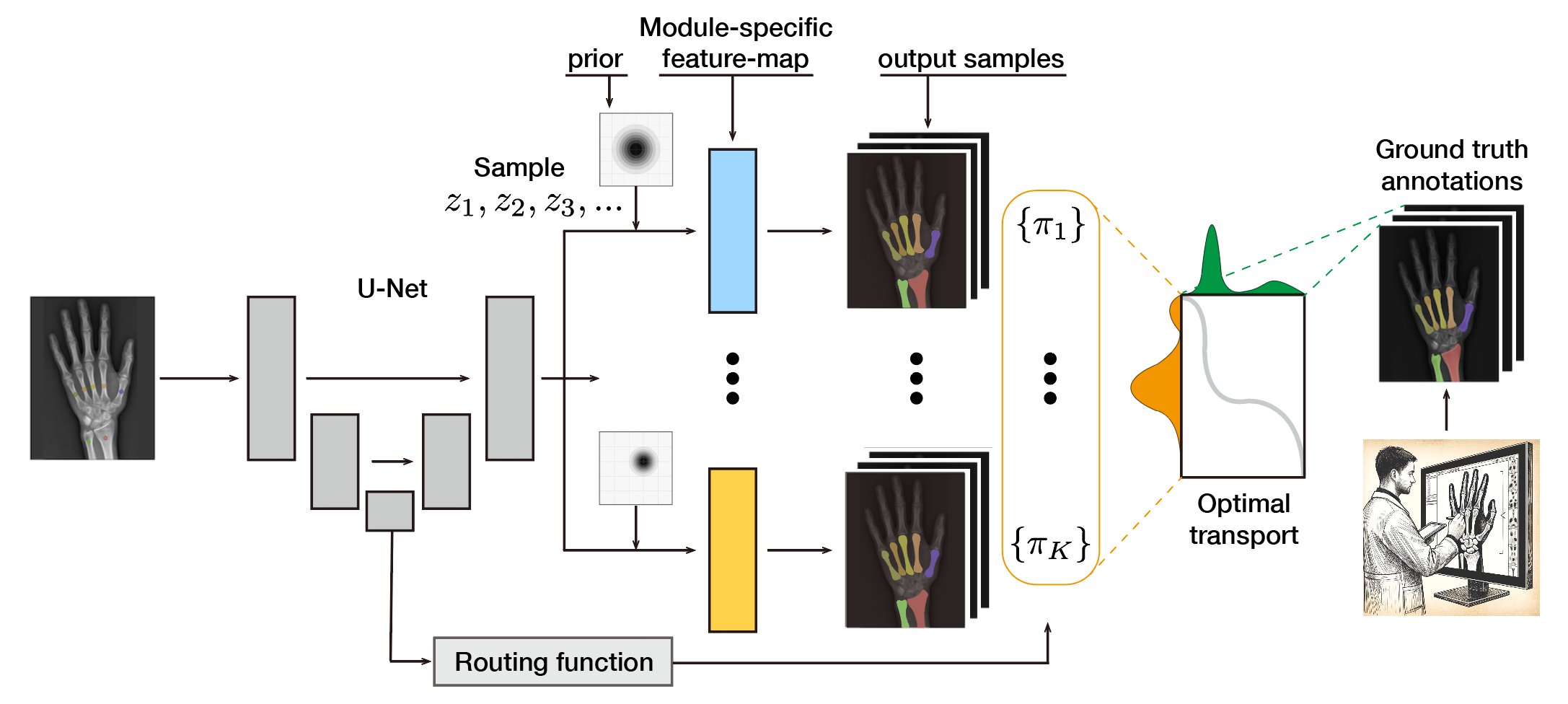}
  \caption{Diagram of the proposed uncertainty-aware segmentation model framework, which features multiple U-Net decoder modules and a routing function that generates segmentation samples along with their respective weights. The model employs an optimal transport-based loss to minimize the gap between the model predictions and the ground truth annotations.}
  \label{fig:seg}
\end{figure}

To facilitate this, we utilize an encoder-decoder structure similar to that of U-Net \citep{ronneberger2015u}, aligning with the model configuration outlined in previous research \citep{hu2019supervised}. A schematic representation of our model is provided in Figure \ref{fig:seg}. We start with a labeled training dataset denoted as $\mathcal{D}=\{(x_i, y_i^{(j)}, \beta_i^{(j)})\}_{i=1:D}^{j=1:M}$, where $x_i\in\mathbb{R}^{H\times W}$ represents the $i$-th sample of a radiographic image, and each image is accompanied by $M$ unique segmentation annotations $y_i^{(j)}\in\{1,...,C\}^{H\times W}$, each associated with a set of probabilities $\beta_i^{(j)}$ that fulfill the condition $\sum_j \beta_i^{(j)}=1$. Here, $C$ denotes the total number of class labels, while $H$ and $W$ indicate the image's height and width, respectively. Our objective is to accurately predict the probability distribution for the segmentation outputs across multiple annotated images.

\subsubsection{Uncertainty-Aware Segmentation with Modularity}
Our segmentation framework is designed around an encoder-decoder architecture, where the encoder function $f_{\text{enc}}(x)$ transforms an input image $x$ into a global feature map $u_e \in\mathbb{R}^{F_1\times H'\times W'}$, and the decoder function $f_{\text{dec}}(u_e)$ maps this feature to a decoded feature space $u_d \in\mathbb{R}^{F_2\times H\times W}$. The decoder output $u_d$ is then integrated with a module-specific noise sampled from an isotropic Gaussian prior, $z_k \sim \mathcal{N}(\mu_k,\sigma_k^2 I)$, to generate the output of the $k$-th segmentation module as $\bar{u}_d^k[:,i,j] = W_s z_k\odot(u_d[:,i,j]+W_b z_k)$. Here, $\mu_k \in \mathbb{R}^L$ represents the mean vector, $I$ denotes an $L\times L$ identity matrix, and $W_s,W_b\in\mathbb{R}^{F_2\times L}$ are matrices employed for the linear transformation of the latent noise, which is then used to scale and bias the decoded feature map, respectively. The final prediction for the $k$-th module, $s_k \in\mathbb{R}^{C\times H\times W}$, is then computed using three $1\times 1$ convolution layers as $s_k=f_{\text{Conv1D}}\left(\bar{u}_d^k\right)$. This setup allows each module to capture a different aspect of aleatoric uncertainty \citep{hullermeier2021aleatoric}, thereby representing the posterior predictive distribution through a collection of weighted samples:
\begin{align}
q_{\theta}(y | x)=\sum_{k=1}^K \pi_k(x) \left(\frac{1}{S} \sum_{j=1}^S \delta(s_k^{(j)})\right) \label{eq1}
\end{align}
This formula assumes that for each $k$-th module, we generate $S$ samples, referred to as $\{s_k^{(j)}\}_{j=1}^S$, by initially sampling an equal number of latent codes $z_k$. We then calculate the average of the module-specific predictions to obtain each module's predictive distribution. These individual distributions are subsequently merged to form the overall predictive distribution $q_\theta$, with the weighting for each module given by $\pi_k(x)$. The determination of these weights is managed by an additional routing function $r(\cdot)$ (see Figure \ref{fig:seg}). This function applies average pooling on the feature map $u_e$ to yield a condensed vector $\bar{u}_e = \text{AvgPool}(u_e)\in\mathbb{R}^{F_1}$, which is then processed through a multi-layer perceptron (MLP) that assigns a probabilistic weight $\pi_k(x)$ to each module's contribution. The complete set of parameters within the model is represented by $\theta$.

\subsubsection{Optimal Transport-based Loss}
We define the learning process of our model as the minimization of a Wasserstein loss \citep{ruschendorf1985wasserstein} between the model prediction $q_\theta(y|x)$ and the actual annotated distribution of the image $p(y|x)$:
\begin{align}
    q_\theta(y|x) &=\sum_{i=1}^N \alpha^{(i)}\delta(s^{(i)})=\sum_{k=1}^K\sum_{j=1}^S\frac{\pi_k(x)}{S}\delta(s_k^{(j)}) \label{eq2}\\
    p(y|x) &=\sum_{j=1}^M\beta^{(j)}\delta(y^{(j)}) \label{eq3}
\end{align}
This is approached as an optimal transport (OT) problem \citep{villani2009optimal,peyre2019computational}, which determines the most cost-efficient way to reallocate the probability mass from $q_\theta(y|x)$ to the true distribution $p(y|x)$, guided by minimizing a cost matrix $\mathcal{C}\in\mathbb{R}^{N\times M}$. For the cost $\mathcal{C}_{ij}$ between pairs $s^{(i)}$ and $y^{(j)}$, we employ the generalized energy distance (GED) \citep{kohl2018probabilistic,baumgartner2019phiseg}, which is often used to assess the alignment of the model predictions with the ground truth labels within the segmentation output space. In contrast to the Kullback-Leibler divergence, OT offers a well-defined distance measure that remains valid even when the compared distributions have differing supports. Based on this loss, our objective function can be written as follows:
\begin{align}
\min_{\theta_s} & \sum_{n=1}^D\sum_{i=1}^N\sum_{j=1}^M\; \mathcal{T}_{i j}^\ast\, \mathcal{C} \left(s_n^{(i)}(x_n;\theta_s), y_n^{(j)}\right) \label{eq1_ot}\\
\text{s.t.} \quad \mathcal{T}^\ast &=\underset{\theta_{\pi},\,\mathcal{T} \in U}{\arg \min } \sum_{i, j} \mathcal{T}_{i j} \mathcal{C}_{i j}\label{eq2_ot}\\
\text{and }\quad U &=\left\{\mathcal{T} \in \mathbb{R}_{+}^{N \times M}: \mathcal{T} \mathbbm{1}_M = \boldsymbol{\alpha}(x_n;\theta_{\pi}), \mathcal{T}^\top \mathbbm{1}_N=\boldsymbol{\beta}\right\} \label{eq3_ot}
\end{align}
where $\mathcal{T}$ is a coupling matrix indicating the transportation plan, $U$ is the set of all possible values of $\mathcal{T}$ which satisfy the marginal constraints, and $\mathbbm{1}$ is a vector of ones. Equations (\ref{eq1_ot}-\ref{eq3_ot}) constitute a bilevel optimization problem, where the inner problem is responsible for identifying the optimal coupling matrix $\mathcal{T}$, while the outer problem aims to refine the model's output distribution within the framework of this coupling. In contrast to the conventional optimal transport problem, the dependency of the marginal probabilities $\mathbf{\alpha}(x_n;\theta_{\pi})$ on the routing function parameters $\theta_{\pi}$ prevents the direct use of backpropagation for optimizing the entire set of parameters $\theta \triangleq \theta_s\cup\theta_{\pi}$. To overcome this challenge, we use a method of imbalanced constraint relaxation \citep{peyre2019computational}, details of which are further explained in appendix \ref{A1}.

\subsection{Multi-Crop Data Augmentation}\label{sec:aug}
\begin{figure}[t]
\includegraphics[width=\linewidth]{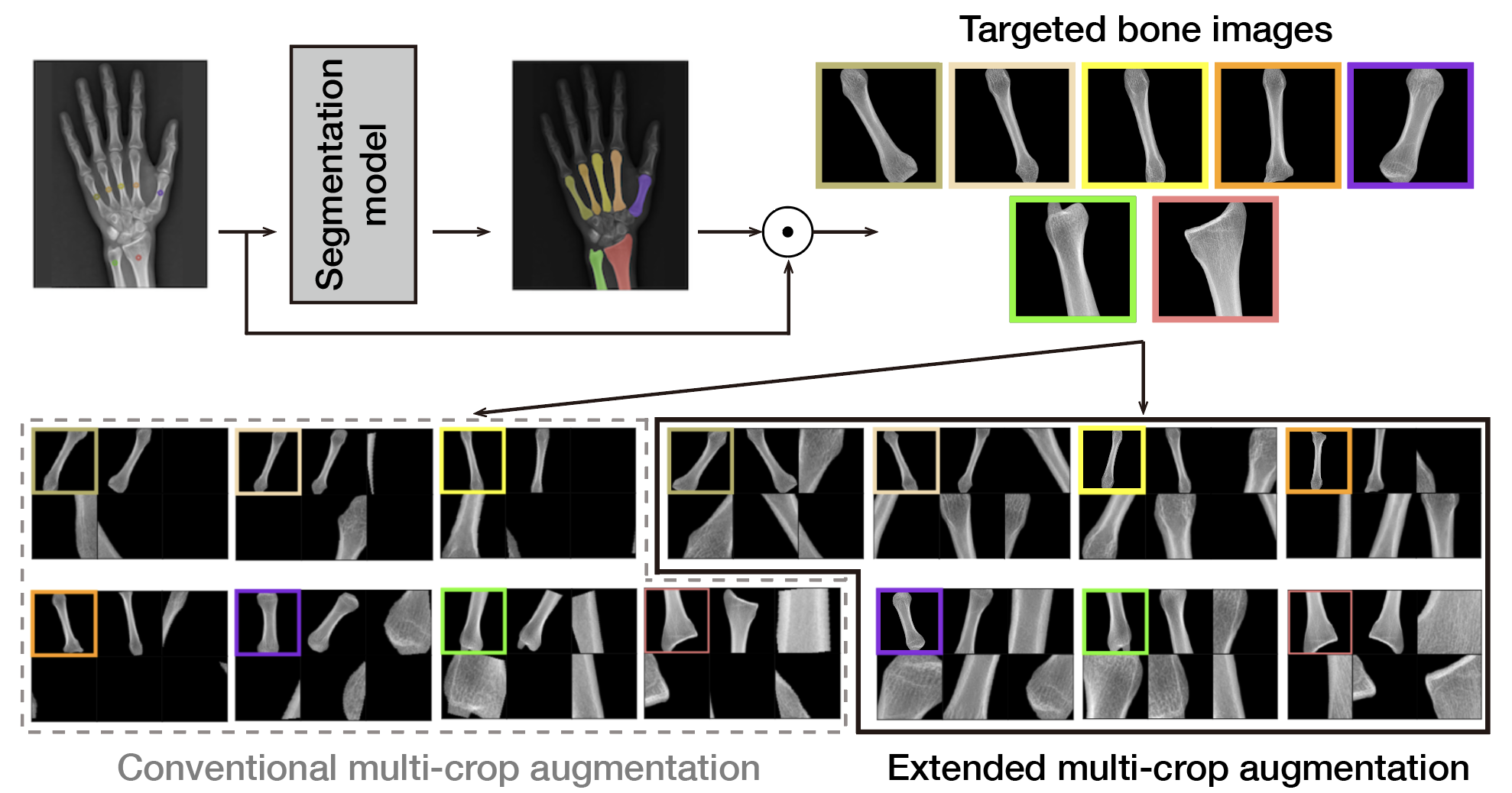}
  \caption{Illustration of the process for extracting targeted bone images from segmentation outputs, and a comparison of two multi-crop augmentation methods. (Top) Seven bone image patches, each outlined by a square in the same color as the corresponding segmentation mask overlaid on the hand X-ray image. (Bottom) Each of the seven image segments undergoes two different multi-crop augmentation strategies: the conventional method (represented by a gray dashed line) and our extended method (indicated by a solid black line). A global view matching the color of the input bone patch and five local views are displayed side by side.}
  \label{fig:aug}
\end{figure}

The resultant output segmentation masks predicted in the previous section are then used to produce targeted bone image segments using a simple Hadamard multiplication. In the end, we derive 7 unique image patches from an original radiograph, corresponding to the ulna, radius, and the five metacarpals (Figure \ref{fig:aug}). It should be noted that there were instances of deviation resulting in fewer than seven target image segments. Such deviations were typically caused by unusual positioning of the hand or the presence of accessories or medical apparatus on the patient's hand.

The bone segments created above exhibit variations in rotational angles (Figure \ref{fig:aug}), primarily due to the inherent anatomical structure of the hand and wrist, as well as the hand position. To equip our model with the ability to reliably process these variations, we implement several data augmentation strategies. These include applying random rotations between $-30$ and $+30$ degrees, horizontal and vertical translations within a range of $-10$ to $+10$\% of the image's width and height, flipping the image randomly in both vertical and horizontal directions, and adjusting the brightness and contrast from $50$ to $150$\% of their original values. Furthermore, we adopt the multi-crop data augmentation method \citep{caron2020unsupervised}, which is advantageous for effective training of pretext tasks as discussed in Section \ref{sec:ssl}. This method samples images of two distinct resolutions: randomly seletected two global views ($224\times 224$) and four local views ($96\times 96$). By using crops of different sizes, the model is exposed to objects at various scales, which enhances its scale invariance.

Conventional multi-crop augmentation, often applied to standard image classification benchmarks or natural images, ensures each cropped segment remains feature-rich. In contrast, our specific bone segments frequently have vast areas of zero pixels (Figure \ref{fig:aug}), a result of the segmentation masking process. This characteristic persists even in the original radiographs before applying the segmentation method due to the sparse anatomical features in hand X-ray images. Consequently, some samples might entirely lack bone content (Figure \ref{fig:aug}). To mitigate this, we tailor the multi-crop strategy to our needs, ensuring that the augmented samples are spatially constrained crops \citep{van2021revisiting,peng2022crafting}. Specifically, our modified method guarantees that both global and local views contain a sufficient amount of feature data, employing iterative rejection sampling to maintain at least a 10\% non-zero pixel presence in each crop. This threshold is set to preserve the integrity of the features, reducing the chance of valuable information being omitted from our samples, as visualized in Figure \ref{fig:aug}. This adjustment secures the inclusion of meaningful content in each cropped segment, optimizing their contribution to the pre-training stage.

\subsection{Contrastive Self-Supervised Pretraining}
\label{sec:ssl}
\begin{figure}[t]
\includegraphics[width=\linewidth]{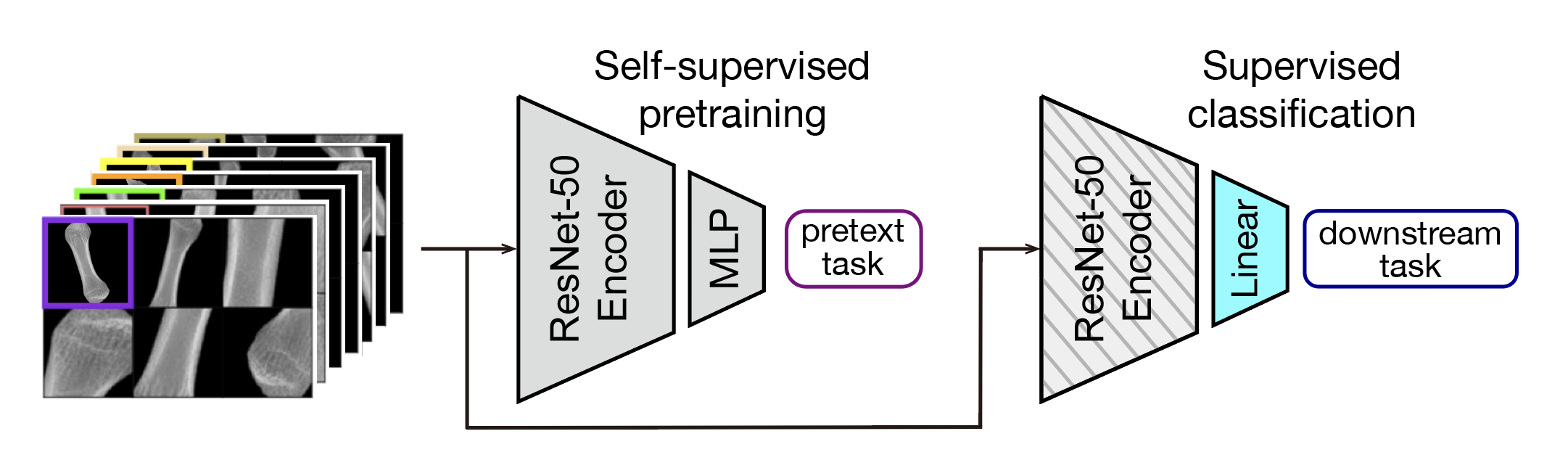}
  \caption{Two-stage training for osteoporosis prediction. In the first stage, augmented samples are processed through an encoder for self-supervised pretraining as part of a pretext task. The backbone classification encoder network (gray trapezoid with slanting lines) is subsequently fixed and repurposed to address the downstream classification task by incorporating a trainable linear layer (cyan trapezoid).}
  \label{fig:ssl}
\end{figure}

Our next step progresses through two separate training phases, as illustrated in Figure \ref{fig:ssl}. The first phase involves a pretext task where we train the ResNet-50 encoder using the augmented views described previously. This phase operates under the premise that transformations applied to an image do not change its semantic content. Consequently, the task focuses on bringing closer the representations of different augmentations of the same image, referred to as a positive pair, in the latent space. This crucial step allows the encoder to focus on the inherent data structures and enables the model to learn robust, generalizable features from the radiographic images without relying on human-provided labels. For the optimization of this process, we employ the LARC optimizer \citep{you2017large}, along with a cosine annealing scheduler \citep{loshchilov2016sgdr} to adjust the learning rates. The entire training procedure is executed with a batch size of 128.

In this phase, we assess four well-known SSL techniques: SimCLR \citep{chen2020simple}, SupCon \citep{khosla2020supervised}, SwAV \citep{caron2020unsupervised}, and VICReg \citep{bardes2021vicreg}, all categorized under contrastive self-supervised methods. They collectively aim to maximize consistency among augmented views within a shared embedding space, ensuring that semantically similar data points are brought closer together. Specifically, SimCLR \citep{chen2020simple} concentrates on contrasting different augmented views of the same instance against other instances, while SupCon \citep{khosla2020supervised} extends this approach by also pulling instances of the same class closer together in the embedding space. SwAV \citep{caron2020unsupervised} introduces a swapped prediction mechanism that bypasses the need for negative pair comparisons, making it computationally more efficient. Meanwhile, VICReg \citep{bardes2021vicreg} stands out by imposing variance, invariance, and covariance regularization to the latent space to ensure the features captured are diverse and not merely reflective of trivial patterns. Each method thus contributes a unique approach to overcoming the challenges of self-supervised learning in its own way. Exploring these effective SSL methods in peripheral radiographic images still remains nascent, making our experimental results potentially useful for clinical applications in other medical domains.

\subsection{Supervised Fine-Tuning}
\label{sec:finetuning}
Following the initial phase, the pre-trained encoder is fine-tuned on a dataset that is explicitly labeled for our classification task. In this downstream phase, the encoder serves as an efficient feature extractor, with only its final linear layer being trained to precisely predict osteoporosis. Importantly, the inputs to this pre-trained encoder are not raw hand X-ray or cropped images, but rather bone segments isolated by the segmentation model. We initially aim to perform subject-wise osteoporosis screening; however, the current training model is not suited for such subject-wise predictions as it outputs multiple sub-binary decisions for different bone segment inputs from each individual. To address this, we maintain the bone segment-wise input approach but revise the evaluation protocol by aggregating multiple sub-decisions to form a final subject-wise decision. Specifically, each sub-decision from the bone segments yields a probability score for each class. The final decision is derived from the average probability across the seven image segments of the ulna, radius, and five metacarpals:
\begin{align}
    \hat{y} = \mathop{\arg\max}_{y_j} \; \langle p(y_j | x_i) \rangle_{i=1:7}
\end{align}
where $p(y_j | x_i)$ represents the probability predicted by the $i$-th bone image segment $x_i$ for class $y_j$, and the final class prediction $\hat{y}$ is determined by the class with the highest average probability. Through this approach, the classifier is trained to uniformly identify osteoporosis across all bone segments during the fine-tuning phase. At the time of evaluation, the assessment criteria are adjusted so that the model's prediction is considered accurate even if only a few bone segments are diagnosed as positive. This adjustment ensures that the model's ability to detect osteoporosis is evaluated based on its overall performance across various segments, rather than requiring a unanimous positive diagnosis from all segments.

\section{Datasets \& Metrics}
\paragraph{Hand X-ray Datasets}
We leverage two primary datasets of hand X-ray images. The first dataset consists of IRB-approved hand radiographic images from 192 individual subjects, featuring a resolution of 2515 × 3588. About 28$\sim$29\% of these subjects are diagnosed with osteoporosis. For precise classification, we correlate our data with DXA scans. DXA measures BMD by assessing the variation in X-ray attenuation across different bone areas, quantifying the concentration of calcium and other minerals in these segments. The derived BMD values are used to compute the T-score, which benchmarks a patient's BMD against the average BMD of a healthy 30-year-old adult. A negative T-score indicates a BMD lower than average, while a positive T-score denotes a higher than average BMD. In our study, this T-score is employed to annotate radiographic images, categorizing an X-ray as indicative of osteoporosis if the T-score falls below -2.5, and as normal otherwise.

Additionally, we use a publicly available dataset \citep{hand-xray_dataset} containing 1154 hand X-ray images, each measuring 1400 × 900. Originally compiled for an object detection benchmark and lacking osteoporosis labels, this dataset is exclusively utilized for training our bone segmentation model and pretraining the backbone classification encoder network. We apply a subject-wise random split of 80\% for training, 10\% for validation, and 10\% for testing, maintaining this ratio consistently across all experiments and different seeds.

To generate the ground truth labels required for training our segmentation model, we leverage the Segment Anything Model (SAM) \citep{kirillov2023segany}, a foundational model for image segmentation. Our initial experiences indicate that SAM necessitates manual interventions, such as placing positive and negative point prompts inside and outside the target bone regions respectively, to produce segmentation masks that match the expert-level precision of radiologists and orthopedists. Our ultimate goal is to develop a fully automated osteoporosis screening model that integrates classification through segmentation without relying on manual segmentation steps. Therefore, we employ SAM as an auxiliary tool for creating ground truth bone segments, guided by expert feedback.

\paragraph{Metrics} 
All test results are obtained from the optimal model identified during training, which is determined by achieving the highest validation macro-F1 score. This measure provides a reliable assessment of model performance, especially under conditions of imbalanced labels \citep{fiorillo2021deepsleepnet}. For evaluation purposes, we employ conventional classification performance metrics, such as precision, recall, F1 score, AUC, and accuracy rate. We report them specifically for the positive label. Particular emphasis is placed on the AUC because of the class imbalance in our dataset. A higher AUC value indicates that the classifier is more effective at distinguishing between positive and negative cases of osteoporosis screening across various classification thresholds.

\section{Experimental Results} 

\begin{table*}[t]
  \caption{Performance comparison of osteoporosis prediction using the proposed model framework on held-out radiographic test images. The table reports the precision, recall, F1-score, AUC, and accuracy for four different methods (SimCLR, SupCon, SwAV, and VICReg), with each metric averaged over three trials initiated with distinct random seeds. The highest performance for each metric is highlighted in bold.}
  \label{tab:exp1}
{
\resizebox{\textwidth}{!}{
\begin{tabular}{c|ccccc}
\hline
       & Precision & Recall & F1 & AUC & Accuracy \\ 
\hline
SimCLR & \textbf{0.71 $\pm$ 0.04} & \textbf{0.66 $\pm$ 0.06} & \textbf{0.68 $\pm$ 0.03} & \textbf{0.85 $\pm$ 0.01} & \textbf{0.81 $\pm$ 0.03} \\
SupCon & 0.61 $\pm$ 0.01 & 0.60 $\pm$ 0.11 & 0.60 $\pm$ 0.06 & 0.79 $\pm$ 0.03 & 0.75 $\pm$ 0.04 \\
SwAV   & 0.54 $\pm$ 0.12 & 0.54 $\pm$ 0.16 & 0.54 $\pm$ 0.13 & 0.68 $\pm$ 0.04 & 0.72 $\pm$ 0.05 \\
VICReg & 0.64 $\pm$ 0.07 & 0.64 $\pm$ 0.07 & 0.64 $\pm$ 0.07 & 0.80 $\pm$ 0.03 & 0.77 $\pm$ 0.03 \\ 
\hline
\end{tabular}
}}
\end{table*}

\subsection{Evaluation on Hand X-ray Images}
The results presented in Table \ref{tab:exp1} compare the performance of four different self-supervised learning models -- SimCLR, SupCon, SwAV, and VICReg -- on the task of osteoporosis prediction using five different metrics previously described. Each model was evaluated across three trials with distinct random seeds, providing an average for each metric. Among these models, SimCLR achieves the highest F1 score of 0.68 $\pm$ 0.03 and an AUC of 0.85 $\pm$ 0.01, indicating strong predictive performance and consistency in ranking the positive class. In contrast, SwAV appears to struggle relative to the other models, with both its F1 and AUC scores the lowest at 0.54 $\pm$ 0.13 and 0.68 $\pm$ 0.04 respectively, suggesting potential challenges in handling our specific dataset or task. VICReg and SupCon, showing a balanced performance across precision, recall, and F1, exhibits a reasonably good AUC, demonstrating effective but not optimal performance.

\begin{figure}[t]
\includegraphics[width=0.9\linewidth]{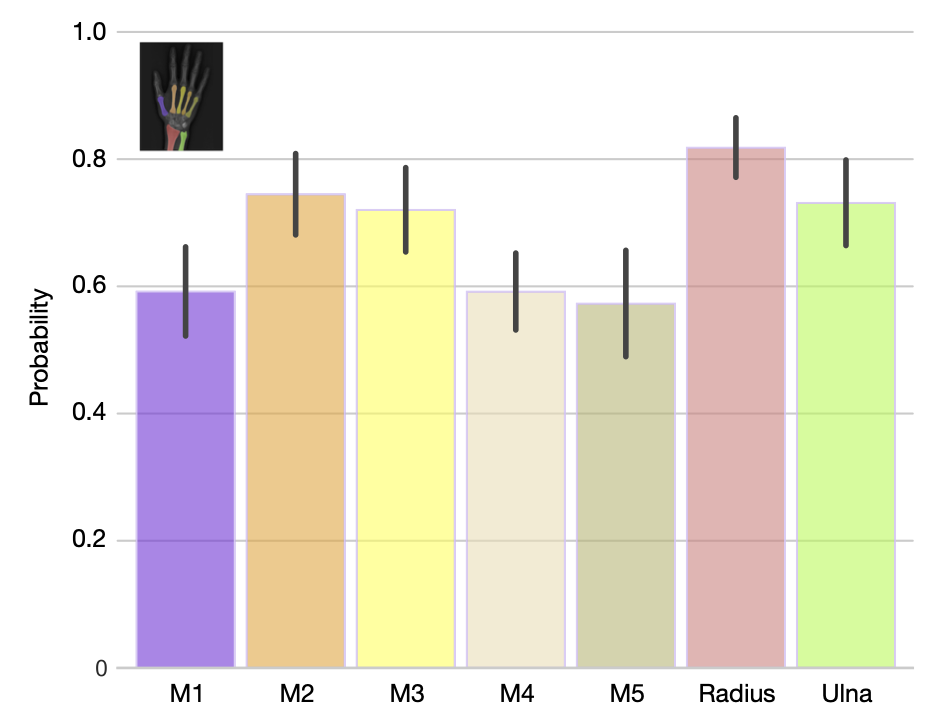}
  \caption{Individual contributions of bone segments to osteoporosis detection. Predicted probabilities for positive labels, illustrating the role of each bone segment, including metacarpals (M1–M5), radius, and ulna, in detecting osteoporosis. Error bars indicate the uncertainty of each segment's prediction. The bar colors correspond to the segmentation mask colors shown in the inset.}
  \label{fig:sub_decision}
\end{figure}

\subsection{Effect of Individual Bones on Osteoporosis Detection}
While the previous experiment demonstrated the model’s accuracy in diagnosing osteoporosis among individuals, we aimed to further investigate the model’s decision-making process by visualizing the predicted probabilities of osteoporosis for each bone involved in the final prediction. We gathered probability values from individual bone images of true positive cases, specifically from our best-performing SimCLR-based classification model. Our analysis indicated that, on average, the 2nd metacarpal, radius, and ulna are the primary contributors to detecting osteoporosis (Figure \ref{fig:sub_decision}). This outcome is in line with studies that recognize the 2MCI as a reliable biomarker, and it corroborates clinical findings that the radius and ulna -- integral parts of the forearm -- are critical in osteoporosis diagnosis.

\subsection{Ablation Studies}
To gain more insights into the features that drive the model’s predictions, we performed an ablation on our proposed model framework. This process involves selectively removing one of the following steps in a series: i) segmenting raw radiological images, ii) applying enhanced multi-crop augmentation to the masked images, iii) pre-training encoders through self-supervised learning, and iv) fine-tuning classifiers in a supervised manner. Initially, to determine whether bone segmentation is essential for osteoporosis prediction, we redesigned an experiment in which the segmentation step is omitted. In this setup, raw X-ray images are directly used as inputs for both the pre-training and fine-tuning stages. Consequently, the final class prediction relies on a single model prediction probability rather than an average of multiple predictions. Despite these changes, pre-training proceeds with the same data augmentation and SSL methods as in previous experiments. To ensure a fair comparison with the original experiment, we maintained the same number of augmented image samples provided to the encoder during pre-training by randomly sampling seven image patches from each hand X-ray image. Table \ref{tab:exp2} displays the prediction results, focusing on three key metrics: F1 score, AUC, and accuracy. Across all models, the omission of bone segmentation results in a noticeable decline in model performance. Specifically, SimCLR experienced the largest decrease in F1 score by 0.12, AUC by 0.07, and accuracy by 0.07, followed closely by VICReg, which showed similar declines in both metrics. These findings underscore the importance of the segmentation step in improving model accuracy and reliability, demonstrating that precise localization of relevant bone features significantly contributes to the performance of osteoporosis screening. Conversely, the results suggest that task-irrelevant features, such as soft tissues including muscles, fat, skin, and nerves -- which absorb X-rays less than bones -- may introduce interference in distinguishing osteoporosis cases.

\begin{table*}[t]
  \caption{Results of osteoporosis prediction after omitting the segmentation step, detailed across three key performance metrics: F1, AUC, and accuracy. The table illustrates the change in model performance (denoted by $\Delta$) when segmentation is excluded, based on averages from three trials with different random seeds. The most significant performance drops are underlined for emphasis.}
  \label{tab:exp2}
{
\resizebox{\textwidth}{!}{
\begin{tabular}{c|cc|cc|cc}
\hline
       & F1 & $\Delta$ & AUC & $\Delta$ & Accuracy & $\Delta$ \\ 
\hline
SimCLR & \textbf{0.54 $\pm$ 0.09} & \underline{-0.12} & \textbf{0.78 $\pm$ 0.04} & \underline{-0.07} & \textbf{0.74 $\pm$ 0.01} & \underline{-0.07} \\
SupCon & 0.51 $\pm$ 0.06 & -0.09 & 0.73 $\pm$ 0.04 & -0.06 & 0.70 $\pm$ 0.06 & -0.05 \\
SwAV   & 0.52 $\pm$ 0.07 & -0.02 & 0.65 $\pm$ 0.06 & -0.03 & 0.71 $\pm$ 0.05 & -0.01 \\
VICReg & \textbf{0.54 $\pm$ 0.07} & -0.10 & 0.74 $\pm$ 0.03 & -0.06 & 0.70 $\pm$ 0.08 & \underline{-0.07} \\ 
\hline
\end{tabular}
}}
\end{table*}

\begin{table*}[t]
  \caption{Impact of removing data augmentation and self-supervised pre-training on osteoporosis prediction. The table presents the F1-score, AUC, and accuracy of the supervised model. The change ($\Delta$) in performance, compared to the baseline model, is also provided for each metric, based on the average results across multiple trials.}
  \label{tab:exp3}
{
\resizebox{\textwidth}{!}{
\begin{tabular}{c|cc|cc|cc}
\hline
       & F1 & $\Delta$ & AUC & $\Delta$ & Accuracy & $\Delta$ \\ 
\hline
Supervised & 0.53 $\pm$ 0.04 & -0.15 & 0.67 $\pm$ 0.03 & -0.18 & 0.72 $\pm$ 0.01 & -0.09 \\ 
\hline
\end{tabular}
}}
\end{table*}

In the second ablation study shown in Table \ref{tab:exp3}, we examined the implications of excluding multi-crop augmentation and SSL pre-training phases while retaining bone segmentation within the model framework. When compared to the optimal results achieved with the SimCLR-based model in Table \ref{tab:exp1}, this current setup demonstrates a considerable drop in performance: the F1 score decreases by 0.15 to 0.53 ± 0.04, AUC by 0.18 to 0.67 ± 0.03, and accuracy by 0.09 to 0.72 ± 0.01. A critical observation is the supervised model's relatively low accuracy of 0.72, which nearly aligns with the proportion of non-osteoporotic patients, suggesting that the model might predominantly predict one class regardless of variations in the input radiographic images. This result emphasizes the limitations of a purely supervised approach in learning meaningful representations for effectively distinguishing osteoporosis from normal cases.

Table \ref{tab:exp4} presents the outcomes of the last ablation study within the original model framework, with a key modification: the implementation of conventional multi-crop augmentation rather than an extended version. The results show a general decline in performance metrics across all models when compared to previous experiments that allowed more bone content in the augmented images. Specifically, SimCLR, which previously showed robust performance, now records a minor dip in every metric. A similar downward trend is observed across all SSL methods. These quantitative reduction suggest that our customized multi-crop augmentation, tailored for hand radiographic images, produces more favorable results, highlighting the significance of pinpointing the sufficient and pertinent bone regions.

\begin{table*}[t]
  \caption{Counter-effect of conventional multi-crop augmentation on osteoporosis screening.}
  \label{tab:exp4}
{
\resizebox{\textwidth}{!}{
\begin{tabular}{c|cc|cc|cc}
\hline
       & F1 & $\Delta$ & AUC & $\Delta$ & Accuracy & $\Delta$ \\ 
\hline
SimCLR & \textbf{0.65 $\pm$ 0.03} & -0.03 & \textbf{0.82 $\pm$ 0.02} & -0.03 & \textbf{0.79 $\pm$ 0.05} & -0.02 \\
SupCon & 0.56 $\pm$ 0.04 & -0.04 & 0.74 $\pm$ 0.04 & -0.05 & 0.72 $\pm$ 0.08 & -0.03 \\
SwAV   & 0.51 $\pm$ 0.09 & -0.03 & 0.66 $\pm$ 0.05 & -0.02 & 0.70 $\pm$ 0.03 & -0.02 \\
VICReg & 0.59 $\pm$ 0.02 & -0.05 & 0.78 $\pm$ 0.03 & -0.02 & 0.74 $\pm$ 0.05 & -0.03 \\ 
\hline
\end{tabular}
}}
\end{table*}

\section{Conclusion}
The prevalence of osteoporosis, a disease that weakens bones and increases the risk of fractures, is rising alongside an aging global population. Diagnosing this condition typically requires analysis of BMD, most commonly through DXA tests, which can be costly and not widely available. Radiographic images, commonly generated in clinical settings, offer a rich, underutilized dataset for detecting osteoporotic changes in bone structure using less invasive methods. In this study, we explored the potential of using hand X-ray images to determine osteoporosis status by introducing the robust segmentation-for-classification framework and leveraging self-supervised pre-training. Our work has a broader impact because this system could greatly reduce the clinician's workload by swiftly pinpointing likely positive cases. Moreover, it could improve patient care by detecting osteoporotic cases that might initially go unnoticed, thus prompting clinicians to reconsider initial assessments and address potential diagnostic biases. Although currently trained on a few hundred samples, we plan to expand our dataset to include tens of thousands of unlabeled radiographs. We are optimistic about the future of this technology, believing it has the potential to lead to more precise and user-friendly osteoporosis screening methods, which we aim to further develop in the future.

\clearpage
\bibliography{refs}

\newpage
\appendix

\section{Constraint Relaxation}
\label{A1}
To tackle the intractability of back-propagation for optimizing the routing parameters $\theta_{\pi}$, we shift the constraint of the predictive marginal from Eq \ref{eq3_ot} to the outer problem outlined in Eq \ref{eq1_relax} as described below:
\begin{align}
\min_{\theta_s,\theta_{\pi}} & \sum_{n=1}^D\sum_{i=1}^N\sum_{j=1}^M\; \mathcal{T}_{i j}^\ast\, \mathcal{C} \left(s_n^{(i)}(x_n;\theta_s), y_n^{(j)}\right) +\lambda D_{\text{KL}}\left( \sum_j \mathcal{T}_{i j}^\ast \,\Vert\, \boldsymbol{\alpha}(x_n;\theta_{\pi}) \right) \label{eq1_relax}\\
\text{s.t.} \quad \mathcal{T}^\ast &=\underset{\mathcal{T} \in U'}{\arg \min } \sum_{i, j} \mathcal{T}_{i j} \mathcal{C}_{i j} \label{eq2_relax}\\
\text{and }\quad U' &=\left\{\mathcal{T} \in \mathbb{R}_{+}^{N \times M}: \mathcal{T} \mathbbm{1}_M \leq \gamma\cdot\mathbbm{1}_N, \mathcal{T}^\top \mathbbm{1}_N=\boldsymbol{\beta}\right\} \label{eq3_relax}
\end{align}
where $\lambda$ serves to balance the conventional segmentation loss (the first term in Eq \ref{eq1_relax}) and the cross-entropy loss (the second term in Eq \ref{eq1_relax}) between the predicted probabilities and a pseudo label generated from the coupling matrix. The range of $\gamma$, from $\frac{1}{N}$ to 1, is strategically chosen to avoid sub-optimal solutions during the initial phases of training. Empirically, we begin with a value $\gamma_0 < 1$ and gradually anneal it to 1 (effectively removing the constraint). The underlying idea is to use the optimally solved matrix $\mathcal{T}^\ast$ as supervision for the routing function in the outer problem outlined in Eq \ref{eq1_relax}.

\end{document}